\documentclass[letterpaper]{article} 
\usepackage{aaai2027}  
\usepackage[hyphens]{url} 
\usepackage{graphicx}  
\usepackage{natbib} 
\usepackage{caption} 
\usepackage{algorithm}
\usepackage{algorithmic}

\usepackage{multirow}
\usepackage{amsmath}
\usepackage{amssymb}
\usepackage{tabularx}

\usepackage{newfloat}
\usepackage{listings}
\DeclareCaptionStyle{ruled}{labelfont=normalfont,labelsep=colon,strut=off} 
\floatstyle{ruled}
\newfloat{listing}{tb}{lst}{}
\floatname{listing}{Listing}

\usepackage{booktabs}

\nocopyright

\title{Towards More Expressive Spoken LLMs: Fine-Grained Intent Benchmarking and Acoustic-Lexical Decoupled Policy Optimization}
\author{
    Xiang Lin\textsuperscript{\rm 1}\footnotemark[1], Tian-Hao Zhang\textsuperscript{\rm 1,2}\thanks{Equal contribution.}, Chunfeng Wang\textsuperscript{\rm 1}, Zhou Pan\textsuperscript{\rm 1}, Kun Zhan\textsuperscript{\rm 1}, Liang Li\textsuperscript{\rm 2}
}
\affiliations{
    \textsuperscript{\rm 1}Li Auto Inc., Beijing, China \\
    \textsuperscript{\rm 2}Tsinghua University, Beijing, China

}

\begin{document}

\maketitle

\begin{abstract}

Spoken emotional dialogue requires a model to understand a user's spoken input and generate a response that is both semantically appropriate and emotionally expressive. This is challenging because communicative intent may be stated explicitly in lexical content or conveyed more implicitly through paralinguistic cues, which can complement or diverge from the words themselves. However, two limitations constrain progress in this area: the scarcity of benchmarks that distinguish these intent expressions, and the lack of reinforcement
learning objectives that jointly account for response quality and emotional expression.
To address the lack of suitable benchmarks, we introduce \textbf{ParaIntent}, a Chinese benchmark comprising 14 intent categories with balanced explicit and implicit samples, together with a multidimensional evaluation protocol covering intent fulfillment, response quality, and emotional expression. For policy optimization, existing approaches either use a shared objective for text and speech or apply reinforcement learning to only one modality, leaving modality-specific learning signals entangled within policy optimization. Motivated by this, we propose \textbf{Acoustic-Lexical Decoupled Policy Optimization (ALPO)}, which computes independent textual and acoustic advantages and routes them to the corresponding text and speech tokens within a unified rollout. Under identical reward functions and training budgets, ALPO improves over standard GRPO on most automatic metrics and achieves the best subjective results among the fine-tuned variants, with particularly clear gains in emotional expressiveness on both the synthetic and human-recorded test sets.
\end{abstract}

\section{Introduction}

\begin{figure}[t!]
    \centering
    \includegraphics[width=0.92\linewidth]{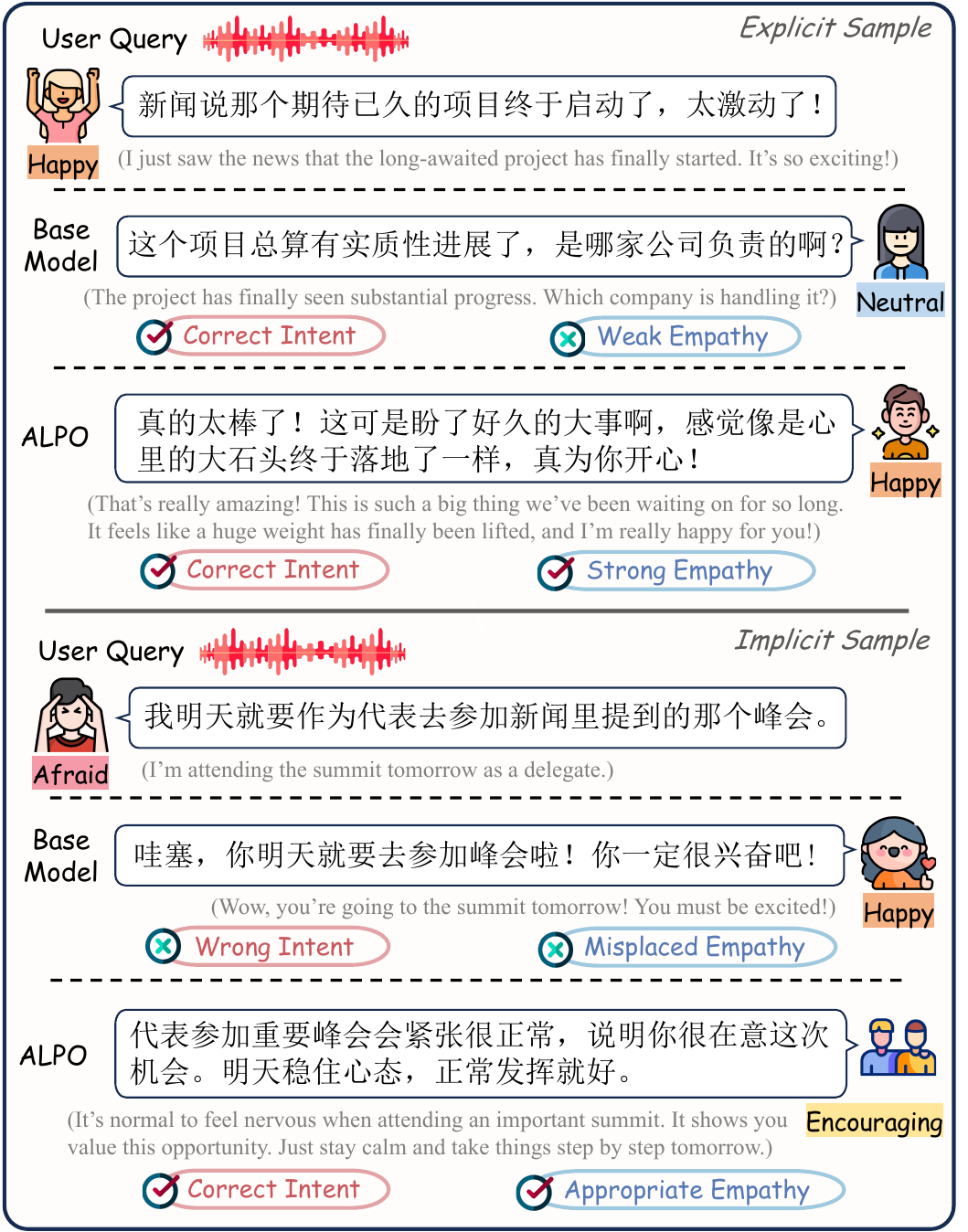}
    \caption{Existing speech dialogue models may produce emotionally limited responses in explicit cases or miss implicit intent; our method generates contextually appropriate, emotionally expressive responses.}
    \label{fig:intro}
\end{figure}

Recent advances in speech large language models have enabled increasingly natural spoken interaction, making it more important for models to faithfully interpret the communicative intent of users~\cite{zeng2024glm,zhang2025mimo}. In spoken dialogue, intent is conveyed not only by what is said, but also by how it is said. Prosody, intonation, and vocal affect can shape utterance interpretation, such that the same lexical content may be understood differently under different vocal delivery styles~\cite{zhou2025echomind}. As illustrated in Figure~\ref{fig:intro}, existing models exhibit distinct weaknesses in explicit and implicit cases. When intent is explicit, they may recover the lexical meaning but produce responses with limited emotional appropriateness. When intent is implicit, they may fail to infer the underlying communicative need conveyed through vocal affect~\cite{wang2026parabridgebridgingparalinguisticperception}. These limitations motivate our study of \emph{spoken emotional dialogue}, which requires models to infer communicative intent from both lexical and paralinguistic speech cues and generate responses coherent in both content and emotion.

Despite its importance, spoken emotional dialogue remains underexplored for two reasons. First, existing benchmarks provide limited coverage of implicit intent. Many datasets align lexical content with vocal affect~\cite{park2024letsrealtalkspoken}, or vary emotion only over semantically neutral text~\cite{yang2025paras2s}, leaving cases with divergent lexical and paralinguistic cues underrepresented. 
Second, existing post-training strategies do not fully account for the distinct roles of text and speech tokens. SFT, commonly used for initialization, optimizes all generated tokens under a single likelihood objective but offers limited control over fine-grained emotional expression~\cite{cheng2025omnichat}. Recent work has therefore explored methods with reinforcement learning (RL)~\cite{kim-etal-2026-aligning}. ParaS2S~\cite{yang2025paras2s} adopts GRPO and assigns the same sequence-level advantage to all response tokens without distinguishing text and speech token types. WavAlign~\cite{chen2026wavalign} uses a hybrid SFT-GRPO objective, applying the GRPO loss only to text tokens while retaining the SFT loss over both text and speech tokens. Thus, ParaS2S lacks speech-specific credit assignment,
whereas WavAlign provides no direct RL signal to speech tokens. These limitations constrain direct optimization of emotional expression and motivate token-specific learning signals for text and speech tokens.

To address these limitations, we introduce \textbf{ParaIntent}, the first Chinese spoken emotional dialogue benchmark to systematically distinguish explicit and implicit intent expressions. It organizes 14 fine-grained intent categories under 4 communicative goals and comprises 140K synthetic training samples, 14K synthetic and 4.2K human-recorded test samples.
ParaIntent further provides a comprehensive evaluation protocol that combines multidimensional automatic assessment of intent fulfillment, response quality, and emotional expression with complementary LLM-based pairwise judgments and human holistic ratings of complete spoken responses.
Building on this benchmark, we propose \textbf{Acoustic-Lexical Decoupled Policy Optimization (ALPO)}, which computes independent textual and acoustic advantages and routes them to the corresponding text and speech tokens within a unified rollout. By removing direct cross-type reward-to-token assignment, ALPO mitigates the trade-off between response quality and emotional expression. Under identical rewards and training budgets, ALPO improves over standard GRPO on most automatic metrics and achieves the best subjective results among the fine-tuned variants, including consistent gains on the human-recorded test set.

Our contributions are summarized as follows:
\begin{itemize}
  \item We introduce ParaIntent, a Chinese benchmark for spoken emotional dialogue spanning 14 intent categories, with balanced explicit and implicit intent samples and a multidimensional evaluation framework covering intent fulfillment, response quality, and emotional expression.
  \item We propose ALPO, a token-level reinforcement learning method that computes independent textual and acoustic advantages and routes them to the corresponding text and speech tokens within a unified policy.
  \item Experiments show that ALPO achieves the strongest overall performance among the fine-tuned variants across automatic and subjective evaluations, with notable gains in emotional expression on both synthetic and human-recorded test sets.
\end{itemize}

\section{Related Work}

\subsection{Spoken Dialogue and Paralinguistic Modeling}

Recent end-to-end speech language models~\cite{wu2025step,ding2025kimi} unify speech understanding and generation and perform well on general spoken interaction tasks. In more controlled settings, StyleTalk~\cite{lin2024advancinglargelanguagemodels} studies responses to identical text spoken in different vocal styles, while OmniChat~\cite{cheng2025omnichat} extends this setup to multi-turn dialogue. These works highlight the value of vocal style but do not explicitly study cases in which intent inference relies more strongly on paralinguistic cues.

\subsection{Benchmarks for Emotional Spoken Dialogue}

Benchmarks such as SD-Eval~\cite{ao2024sd}, EchoMind~\cite{zhou2025echomind}, and CP-Bench~\cite{wang-etal-2025-benchmarking-contextual} assess paralinguistic or contextual speech understanding, while DeepDialogue~\cite{koudounas2025deepdialogue} targets emotionally expressive dialogue. However, they generally assume aligned lexical and vocal-affective cues or do not explicitly model dialogue intent. URO-Bench~\cite{yan2025uro} covers diverse spoken dialogue tasks but lacks implicit-intent scenarios. To our knowledge, no existing benchmark spans fine-grained explicit and implicit intents in Chinese spoken emotional dialogue.

\subsection{Preference Optimization for Speech}
Preference-based methods such as DPO~\cite{rafailov2023direct} and GRPO~\cite{shao2024deepseekmath} are increasingly used to improve generative models beyond supervised fine-tuning. In speech-related tasks, Emo-DPO~\cite{gao2025emo} applies DPO to emotional TTS, while EmotionThinker~\cite{wang2026emotionthinker} uses GRPO with task-specific rewards for speech emotion recognition. For spoken dialogue, ParaS2S~\cite{yang2025paras2s} applies a shared reward to all generated tokens, whereas WavAlign~\cite{chen2026wavalign} restricts reinforcement learning to text tokens and retains SFT for speech generation. ALPO instead computes independent textual and acoustic
advantages and routes them to the corresponding text and speech tokens within a unified rollout.

\begin{figure*}[t!]
    \centering
    \includegraphics[width=1.0\linewidth]{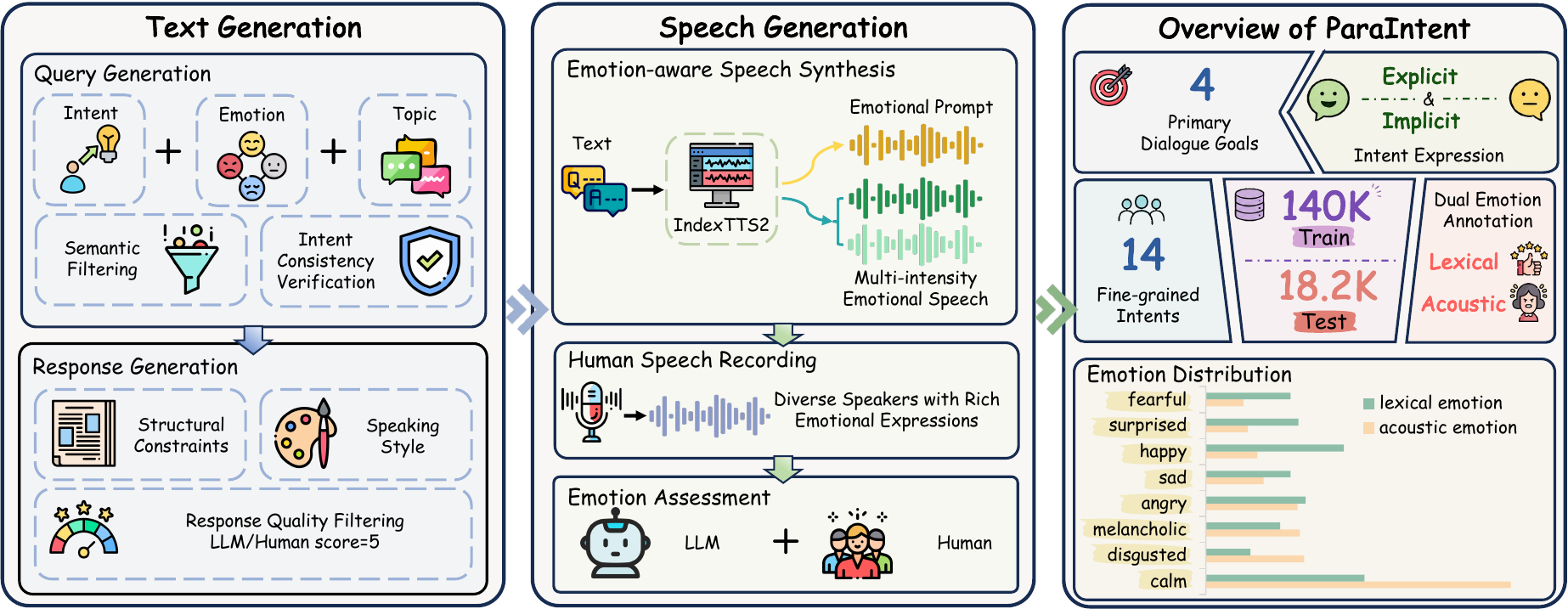}
    \caption{Overview of ParaIntent. (Left) Text generation pipeline: user queries and response texts are produced by GLM-5 under intent-conditioned policies with quality filtering. (Middle) Speech generation pipeline: responses are synthesized with controlled emotion using IndexTTS2, with emotion-intensity variants as preference pairs, alongside a human-recorded test set. (Right) Benchmark overview: 14 intent categories under four communicative goals, with balanced explicit and implicit samples across 140K training and 18.2K test instances.}
    \label{fig:dataset}
\end{figure*}

\section{ParaIntent}
\label{sec:paraintent}

\subsection{Task Definition and Intent Taxonomy}
We formalize spoken emotional dialogue as a single-turn task: given a user speech input $x$, the model infers the underlying intent $i$ from both lexical content and paralinguistic cues and generates a spoken response $y$ appropriate in both content and emotional expression. ParaIntent instantiates this task across explicit and implicit expressions, enabling evaluation of both directly expressed intent and intent inference that relies more strongly on vocal affect.

To organize the task space, we develop a two-level intent taxonomy. The top level comprises 4 broad communicative goals: adversarial interaction, distress support, functional assistance, and social expression. These are further divided into 14 fine-grained intent categories, each associated with a specific response objective and target response emotion. Complete intent definitions, response policies, and representative explicit and implicit examples are provided in the appendix.

ParaIntent balances explicit and implicit samples for each intent.
Explicit intent is largely recoverable from lexical content, with vocal affect providing consistent supporting cues. In implicit cases, intent is less explicit lexically and relies
more on paralinguistic cues, though not on acoustics alone.
Separate evaluation of the two subsets reveals how models handle directly expressed intent and intent conveyed through the interaction of lexical and paralinguistic information.

\subsection{Dataset Construction}
ParaIntent contains 140K TTS-based training samples, 14K TTS-based test samples, and 4.2K human-recorded test samples, as illustrated in Figure~\ref{fig:dataset}. We first use GLM-5~\cite{zeng2026glm} to generate user queries conditioned on intent descriptions, target response emotions, explicit/implicit intent labels, and diverse topics, together with annotations for lexical emotion and vocal affect. 
Training and test queries are independently generated and
deduplicated across splits.
Based on these queries, we further generate response texts using intent information and response policy specifications, followed by automatic and manual quality control.

For speech construction, we use IndexTTS2~\cite{zhou2026indextts2} to synthesize emotionally controlled speech with diverse speaking styles. For each response, we synthesize two variants with different
emotion intensities as chosen and rejected samples for preference
optimization. All samples are further verified for consistency with their target emotion labels. To assess the human-recorded condition, we also build a 4.2K test set with queries and paired reference responses recorded
by five trained native Chinese speakers under a controlled protocol\footnote{The recordings were collected specifically for this study by duly authorized personnel and contain no private information.
}, covering all 14 intent categories.

\begin{figure*}[t!]
    \centering
    \includegraphics[width=0.98\linewidth]{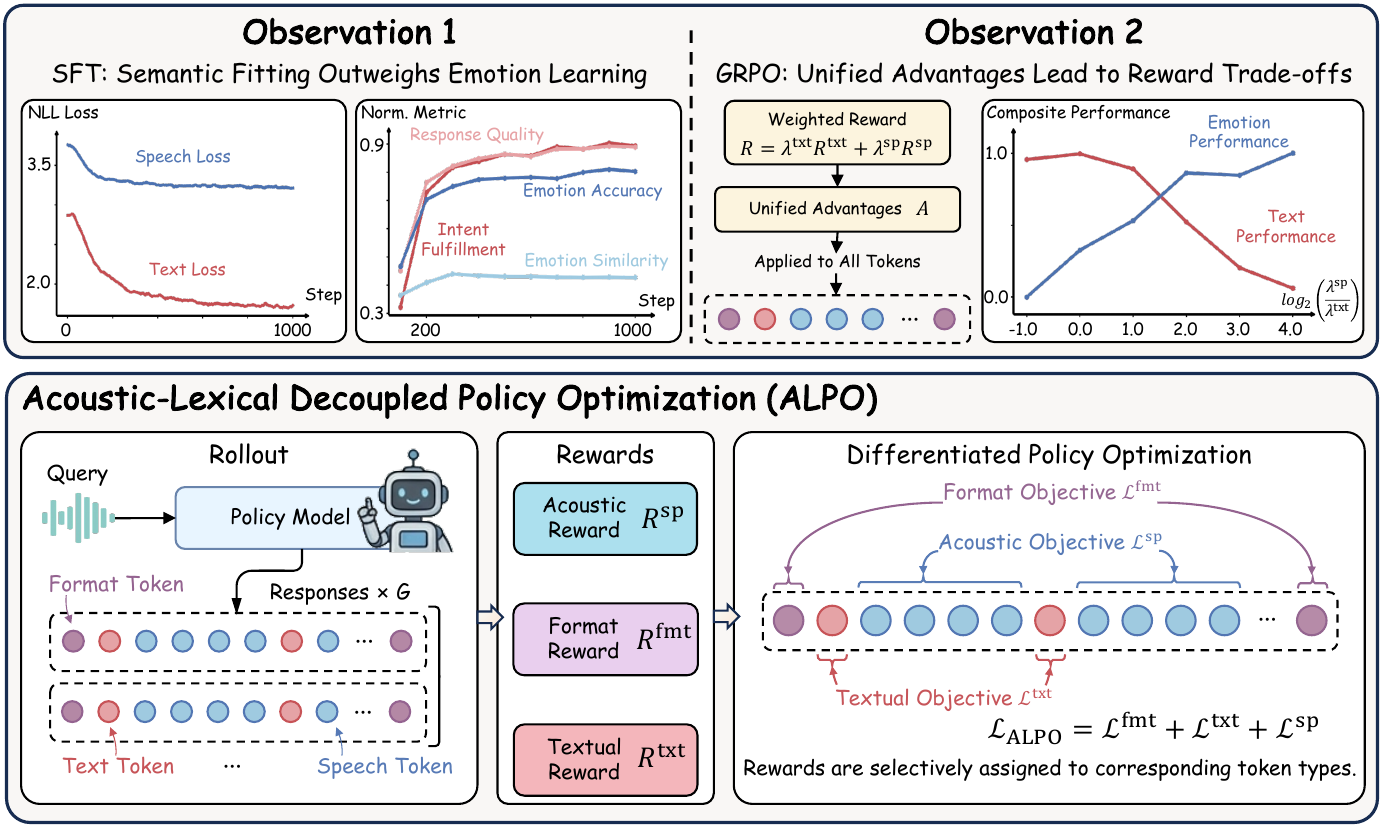}
    \caption{
    ALPO motivation and method. (Top-left) SFT text-token loss decreases faster than speech-token loss, while ES plateaus early. (Top-right) For GRPO, $(\lambda^{\mathrm{txt}},\lambda^{\mathrm{sp}})\in\{(2,1),(1,1),(1,2),(1,4),(1,8),(1,16)\}$. Text and Emotion Performance average IF/RQ and EA/ES, respectively. Increasing the relative acoustic-reward weight improves emotion performance but reduces text performance. (Bottom) ALPO routes separate format, textual, and acoustic advantages to format, text, and speech tokens.
    }
    \label{fig:method}
\end{figure*}

\subsection{Evaluation Protocol}

We use four 0--100 automatic metrics. GLM-5~\cite{zeng2026glm} scores response
text for Intent Fulfillment~(IF), which measures whether the
response addresses the user's intent, while a held-out
DeepSeek-V4-Pro~\cite{deepseekai2026deepseekv4} evaluator scores Response Quality~(RQ),
covering content, linguistic tone appropriateness, and
naturalness. DeepSeek-V4-Pro is not involved in data
construction, reward computation, or training.
MiDashengLM~\cite{dinkel2025midashenglm} label agreement yields speech
Emotion Accuracy~(EA). Emotion Similarity~(ES) uses pretrained
wav2vec2~\cite{baevski2020wav2vec} emotion attributes to
combine similarity to paired reference audio with consistency
against target-emotion statistics. ES remains optimization-aligned
with the acoustic reward, whereas the other metrics provide
held-out or complementary evaluation. 
We additionally validate
ES against emotion-specific human judgments. 
ES correlates with human emotion ratings
(Spearman $\rho=0.681$) and is more sensitive to emotion
mismatch than to controlled speaker variation. We use
speaker-matched references whenever possible; full results
are provided in the appendix.

For subjective evaluation, Gemini-3.1-Pro~\cite{google2026gemini31pro} performs pairwise comparisons of textual quality, emotional expression, and overall preference on spoken responses; human listeners provide holistic ratings.
\section{Method}
\label{sec:method}

\subsection{Preliminary}

Following the task definition above, the model takes a user speech input $x$, infers the underlying intent $i$, and generates a spoken response $y$. We represent $y$ as a single interleaved sequence, where $y^{\text{txt}}$ and $y^{\text{sp}}$ denote the text-token and speech-token subsequences, respectively.
$y^{\text{txt}}$ primarily carries textual content and intent information, while $y^{\text{sp}}$ governs emotional and prosodic realization. The training objective is therefore to jointly improve
textual response quality with respect to user intent and
emotional expression in speech with respect to the target
emotion.

SFT minimizes the negative log-likelihood of the full reference response under teacher forcing:
\begin{equation}
\mathcal{L}_{\mathrm{SFT}}(\theta)
= -\mathbb{E}_{(x,\bar{y})\sim\mathcal{D}_{\mathrm{sft}}}
\sum_{t=1}^{|\bar{y}|}
\log \pi_\theta\!\left(\bar{y}_t \mid x, \bar{y}_{<t}\right),
\end{equation}
where $\bar{y}$ is the full interleaved reference response including both text and speech tokens. SFT provides a strong initialization but does not directly target end-task objectives such as intent fulfillment and emotional expressiveness.

GRPO~\cite{shao2024deepseekmath} samples $G$ candidate responses $\{y_g\}_{g=1}^{G}$ from the old policy $\pi_{\theta_{\mathrm{old}}}$, computes a sequence-level reward $R_g$ for each, and estimates a group-normalized advantage $A_g = \frac{R_g - \mu}{\sigma}$.  The training objective is:
\begin{equation}
\begin{aligned}
\mathcal{L}_{\mathrm{GRPO}}(\theta)=
- \mathbb{E}\,\frac{1}{G}\sum_{g=1}^{G}\frac{1}{|y_g|}\sum_{t=1}^{|y_g|}
\min(
r_{g,t}(\theta)\, A_g,\qquad \quad \\
\mathrm{clip}\!\left(r_{g,t}(\theta),\,1-\epsilon,\,1+\epsilon\right) A_g)
+ \beta\, \mathbb{E}\!\left[\mathrm{KL}\!\left(\pi_\theta \,\|\, \pi_{\mathrm{ref}}\right)\right],
\end{aligned}
\end{equation}
where $r_{g,t}(\theta)=\frac{\pi_\theta\!\left(y_{g,t} \mid x, y_{g,<t}\right)}{\pi_{\theta_{\mathrm{old}}}\!\left(y_{g,t} \mid x, y_{g,<t}\right)}$ is the token-level probability ratio. We set $\beta=0$ and omit the KL penalty, following recent practice~\cite{yu2026dapo}.

\subsection{Limitations of Standard Training Methods}
\label{sec:limitations}

\paragraph{SFT.}

SFT provides a reliable initialization but uniformly optimizes
text and speech tokens despite their distinct roles. For the same response text $y^{\text{txt}}$ and target emotion $e^\star$, multiple speech realizations may be appropriate, whereas single-reference teacher forcing rewards only one. Figure~\ref{fig:method} (Top-left) shows that normalized text-token loss decreases much faster than speech-token loss, suggesting that textual response patterns are easier to learn from a single reference than fine-grained emotional expression. Accordingly, IF and EA improve steadily, while ES plateaus early and slightly declines. Single-reference supervision therefore learns \emph{what to say} effectively, but is less suited to refining \emph{how to say it}.

\paragraph{GRPO.}
GRPO removes the single-reference constraint and enables reward-driven optimization, making it better suited to expressive generation. However, standard GRPO combines textual response quality and emotional expression into a single sequence-level reward:
\begin{equation}
    R = \lambda^{\mathrm{txt}} R^{\mathrm{txt}} + \lambda^{\mathrm{sp}} R^{\mathrm{sp}}.
\end{equation}
A format reward with a fixed coefficient of 1 is also included in all runs but omitted here, since the analysis focuses on the trade-off between textual and acoustic rewards.
The resulting shared advantage $A$ is applied uniformly to all token positions, without distinguishing which reward should guide text or speech tokens. Its induced credit structure can be intuitively decomposed as
\begin{equation}
g \approx
g^{\text{txt}\to\text{txt}} + g^{\text{sp}\to\text{sp}} +
\underbrace{g^{\text{txt}\to\text{sp}} +
g^{\text{sp}\to\text{txt}}}_{\text{cross-type credit assignment}},
\end{equation}
where $g^{A\to B}$ denotes the contribution from reward type $A$ to token type $B$. This is an intuitive characterization, not a strict analytical decomposition, of how a shared advantage induces cross-type credit assignment.

Figure~\ref{fig:method} (Top-right) shows that increasing $\lambda^{\mathrm{sp}}$ improves Emotion Performance but reduces Text Performance, with the reverse trend for $\lambda^{\mathrm{txt}}$; reward weighting alone does not resolve this trade-off. In Figure~\ref{fig:compare} (a), the gradient norm ratio $\frac{\lVert g^{\mathrm{sp}}\rVert}{\lVert g^{\mathrm{txt}}\rVert}$ remains below 1.0 throughout GRPO training, indicating weaker speech-side magnitudes.

In summary, SFT limits expressive diversity through single-reference supervision, while standard GRPO introduces reward-driven optimization but retains mismatched credit assignment across text and speech tokens. This motivates ALPO, which routes independent textual and acoustic advantages to their corresponding token subsets.

\subsection{ALPO}

To address the credit assignment mismatch in standard GRPO, we propose \textbf{A}coustic-\textbf{L}exical Decoupled \textbf{P}olicy \textbf{O}ptimization (ALPO). As illustrated in Figure~\ref{fig:method} (Bottom), ALPO replaces the shared sequence-level advantage with separate advantages for different token types.

\paragraph{Token partition.}
The generated response contains three token types: \emph{format tokens}, which maintain the output structure (e.g., \texttt{<tts\_start>}); \emph{text tokens}, which encode the response content; and \emph{speech tokens}, which represent the acoustic realization. These types occupy non-overlapping vocabulary ranges and are each supervised by a reward for their primary function. Although speech generation is autoregressively conditioned on the preceding text, textual and acoustic rewards evaluate complementary aspects of the response.

\paragraph{Reward design.}
ALPO uses three rewards:

\begin{itemize}
    \item \textbf{Format Reward $R^{\mathrm{fmt}}$} is 1 if the output format is correct, and 0 otherwise.
    \item \textbf{Textual Reward $R^{\mathrm{txt}}$} uses GLM-5 to assess the generated text for content quality, tone appropriateness, and conversational naturalness, normalized to $[0,1]$.
    \item \textbf{Acoustic Reward $R^{\mathrm{sp}}$} combines the emotion similarity of the generated speech to the paired reference and the target-emotion prototype, yielding a score in $[0,1]$.
\end{itemize}

The GLM-5 textual reward targets the same quality dimensions
as RQ, while the acoustic reward directly corresponds to ES.
GRPO and ALPO use identical rewards, isolating the policy
optimization strategy.

\begin{table*}[t]
\centering
\small 
\renewcommand{\arraystretch}{1.05}
\begin{tabularx}{\textwidth}{l *{8}{>{\centering\arraybackslash}X}}
\toprule
\multirow{2}{*}{Model}
& \multicolumn{2}{c}{Intent Fulfillment}
& \multicolumn{2}{c}{Response Quality}
& \multicolumn{2}{c}{Emotion Accuracy}
& \multicolumn{2}{c}{Emotion Similarity} \\
\cmidrule(lr){2-3} \cmidrule(lr){4-5} \cmidrule(lr){6-7} \cmidrule(lr){8-9}
& Syn. & Human & Syn. & Human & Syn. & Human & Syn. & Human \\
\midrule
\multicolumn{9}{l}{\textcolor{gray}{\textit{Open-source Models}}} \\
GLM-4-Voice~\cite{zeng2024glm}        & 33.44 & 30.96 & 45.18 & 41.55 & 40.24 & 35.78 & 43.45 & 42.88 \\
MiMo-Audio~\cite{zhang2025mimo}         & 47.95 & 41.76 & 52.91 & 50.67 & 48.39 & 43.33 & 43.64 & 42.91 \\
Qwen2.5-Omni~\cite{xu2025qwen25omnitechnicalreport}       & 32.78 & 28.64 & 35.68 & 35.12 & 45.94 & 39.89 & 42.21 & 41.27 \\
Qwen3-Omni~\cite{xu2025qwen3}         & 52.27 & 52.00 & 58.65 & 53.81 & 51.18 & 34.11 & 44.22 & 42.38 \\
\midrule
\multicolumn{9}{l}{\textcolor{gray}{\textit{Step-Audio-2-mini Fine-tuned Variants}}} \\
Base~\cite{wu2025step} & 41.48 & 37.17 & 45.87 & 40.04 & 57.88 & 45.87 & 42.23 & 41.88 \\
SFT    & 91.74 & 79.11 & 84.02 & 73.81 & 88.14 & 66.22 & 42.99 & 42.39 \\
DPO    & 91.28 & \textbf{82.39} & 83.49 & 72.37 & \textbf{89.15} & \underline{71.33} & 45.29 & 44.66 \\
GRPO$^\dagger$   & \underline{91.99} & 79.67 & \underline{84.60} & \underline{74.43} & 87.35 & 69.33 & \underline{49.37} & \underline{48.73} \\
Hybrid SFT-GRPO$^\ddagger$ & 87.60  & 77.44 & 80.40 & 72.85 & 72.86 & 62.89 & 45.35  & 44.27 \\
ALPO~(Ours) & \textbf{92.28} & \underline{81.61} & \textbf{85.51} & \textbf{77.84} & \underline{88.72} & \textbf{71.78} & \textbf{51.92} & \textbf{51.25} \\
\bottomrule
\end{tabularx}
\caption{
Automatic results on the synthetic and human-recorded ParaIntent test sets. 
Bold = best, underlined = second-best fine-tuned variant.
GRPO$^\dagger$ follows ParaS2S~\cite{yang2025paras2s}; Hybrid SFT-GRPO$^\ddagger$ follows WavAlign~\cite{chen2026wavalign}.}
\label{tab:main_results}
\end{table*}

\paragraph{Token-Specific Advantage Computation and Routing.}
ALPO independently group-normalizes each reward and applies its advantage only to the corresponding tokens. For $k\in\{\mathrm{fmt},\mathrm{txt},\mathrm{sp}\}$:
\begin{equation}
A_g^{k} = \frac{R_g^{k} - \mu_{k}}{\sigma_{k}+\varepsilon},
\end{equation}
where $\mu_k$ and $\sigma_k$ are the group-wise mean and standard deviation for reward type $k$. Let $I_k(y_g)$ index tokens of type $k$ in $y_g$. The per-type loss is:
\begin{equation}
\begin{aligned}
\mathcal{L}^{k}(\theta)
=& - \mathbb{E}\,\frac{1}{G}\sum_{g=1}^{G}\frac{1}{|I_k(y_g)|}\sum_{t\in I_k(y_g)}
\min\!\Bigl(
r_{g,t}(\theta)\, A_g^{k}, \\
& \qquad \mathrm{clip}\! \bigl(r_{g,t}(\theta),1{-}\epsilon,1{+}\epsilon\bigr)\, A_g^{k}
\Bigr),
\end{aligned}
\end{equation}
and the final objective is:
\begin{equation}
\mathcal{L}_{\mathrm{ALPO}}(\theta)=
\mathcal{L}^{\mathrm{fmt}}(\theta)
+\mathcal{L}^{\mathrm{txt}}(\theta)
+\mathcal{L}^{\mathrm{sp}}(\theta).
\end{equation}

Independent advantage normalization prevents differences in
reward scale and variance from dominating a shared
advantage, while preserving reward-specific directions
within each rollout. For example, a response may receive a
positive acoustic advantage but a negative textual
advantage when its speech is emotionally appropriate but
its response text is poor. Concurrent methods such as
GDPO~\cite{liu2026gdpogrouprewarddecouplednormalization}
and DVAO~\cite{jiang2026dvaodynamicvarianceadaptiveadvantage}
also normalize rewards independently, but apply every
advantage to all token positions. ALPO further routes each
advantage only to its corresponding token subset, so
textual rewards do not directly weight speech-token updates,
and acoustic rewards do not directly weight text-token
updates. 
This removes direct cross-type reward-to-token
assignment while retaining all token types within a unified
policy. 
By keeping textual and acoustic objectives separate, ALPO
avoids combining $R^{\mathrm{txt}}$ and
$R^{\mathrm{sp}}$ into a single weighted advantage. The
following experiments evaluate this design through reward
trajectories, downstream performance, and gradient
statistics.

\section{Experiment}

\subsection{Experimental Setup}

All experiments are conducted on ParaIntent using the 140K synthetic training set and are evaluated on the 14K synthetic and 4.2K human-recorded test sets. We compare representative open-source spoken dialogue models. Using Step-Audio-2-mini~\cite{wu2025step} as the backbone, we evaluate Base, SFT, DPO, GRPO, Hybrid SFT-GRPO, and ALPO. GRPO follows the shared-advantage design of ParaS2S~\cite{yang2025paras2s}, while the hybrid method follows WavAlign~\cite{chen2026wavalign} by applying reinforcement learning only to text generation. 
We further compare independent per-reward advantage normalization without token-specific routing, following the normalization principle of GDPO.
DVAO additionally introduces variance-adaptive advantage weighting and is therefore discussed as a related concurrent approach rather than reproduced exactly.
GRPO and ALPO use identical rewards, and all backbone-based methods use the same data splits and training budget. Each trainable configuration is run once, and all reported results use its final checkpoint under the shared training budget, without validation-based checkpoint selection.

Automatic evaluation measures IF and RQ for generated text
and EA and ES for generated speech. IF is scored by GLM-5
but is not used as a training reward, whereas RQ is evaluated
by held-out DeepSeek-V4-Pro. ES remains reward-aligned, and
IF and EA provide complementary task-level evaluation.
For subjective evaluation, Gemini-3.1-Pro conducts
round-robin pairwise comparisons among Base, SFT, DPO,
GRPO, and ALPO on the synthetic test set. We report average
win rates for textual quality, emotional expression, and
overall preference. Five native Chinese annotators separately
provide 0--5 holistic ratings for the same systems on 50
synthetic and 50 human-recorded test instances.

\subsection{Main Results}

Table~\ref{tab:main_results} reports the main results on
ParaIntent. SFT, DPO, GRPO, and ALPO substantially
outperform the base model and existing open-source systems on
most metrics, demonstrating the importance of task-specific
adaptation. The WavAlign-style Hybrid SFT-GRPO variant
improves ES over SFT on both test sets, but yields lower IF,
RQ, and EA, revealing a trade-off between its gains in
emotional similarity and performance on the other evaluation
dimensions.

Among the fine-tuned variants, ALPO ranks first on six of the
eight metrics and second on the remaining two.
Compared with GRPO, ALPO
improves all four metrics on both test sets. In particular, ES
increases from 49.37 to 51.92 on the synthetic test set and
from 48.73 to 51.25 on the human-recorded test set. On human-recorded data, ALPO achieves the best RQ, EA, and
ES while ranking second in IF, whereas DPO obtains the
highest human-recorded IF and synthetic EA. Overall, ALPO achieves a more favorable balance between
textual response quality and emotional expression than the
evaluated ParaS2S-style shared-advantage and WavAlign-style
text-only RL baselines.

\begin{table}[t]
\centering
\small 
\renewcommand{\arraystretch}{1.0}
\begin{tabularx}{\columnwidth}{l *{5}{>{\centering\arraybackslash}X}}
\toprule
\multirow{2}{*}{Model}
& \multicolumn{3}{c}{Gemini Pairwise Win Rate}
& \multicolumn{2}{c}{Human Rating} \\
\cmidrule(lr){2-4}
\cmidrule(lr){5-6}
& Text & Emotion & Overall & Syn. & Record \\
\midrule
Base & 12.44 & 12.47 & 11.08 & 0.66 & 0.65 \\
SFT  & 58.16 & 55.88 & 56.28 & 3.90 & 3.54 \\
DPO  & 57.26 & 58.17 & 58.17 & 4.15 & 4.05 \\
GRPO & 61.02 & 61.02 & 61.51 & 4.32 & 4.24 \\
ALPO & 61.13 & 62.45 & 62.96 & 4.78 & 4.72 \\
\bottomrule
\end{tabularx}
\caption{
Subjective evaluation. Gemini-3.1-Pro pairwise win rates
are evaluated on the synthetic test set. Human holistic
ratings are evaluated on 50 synthetic and 50 human-recorded
test instances. Higher is better.
}
\label{tab:subjective_syn}
\end{table}

Table~\ref{tab:subjective_syn} reports the subjective results.
ALPO achieves the highest Gemini pairwise win rates for
textual quality, emotional expression, and overall preference,
together with the highest human ratings on both
conditions. Compared with GRPO, it improves the
Emotion/Overall win rates from 61.02/61.51 to 62.45/62.96,
while increasing human ratings from 4.32/4.24 to 4.78/4.72
on synthetic/human-recorded subsets. These results provide
complementary perceptual evidence that ALPO boosts
response quality and emotional expression beyond the
optimization-aligned automatic metrics.

\begin{figure*}[t!]
    \centering
    \includegraphics[width=1.0\linewidth]{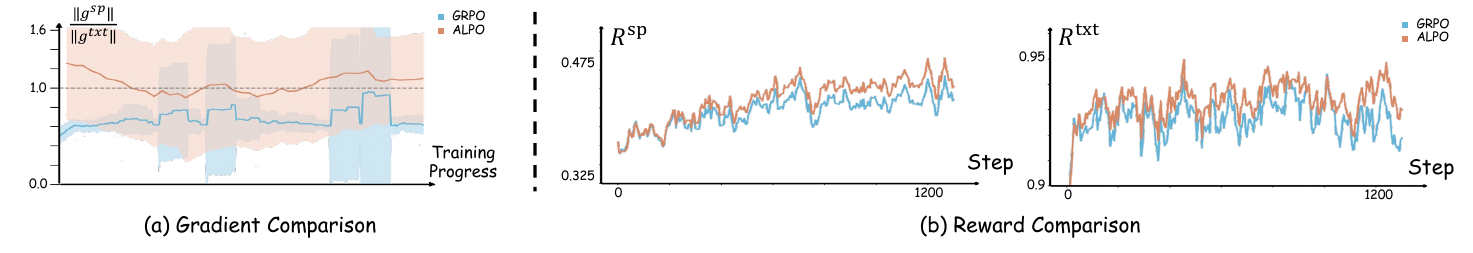}
    \caption{Training dynamics of GRPO and ALPO. (a) Gradient norm ratio: ALPO produces more balanced text- and speech-side gradient magnitudes. (b) Reward curves for $R^{\mathrm{txt}}$ and $R^{\mathrm{sp}}$: ALPO reaches higher values on both optimization objectives.}
    \label{fig:compare}
\end{figure*}

\subsection{Explicit vs.\ Implicit Intent Analysis}

Table~\ref{tab:explicit_implicit} reports explicit and implicit
results. All models achieve higher IF on explicit samples, indicating that implicit intent is harder. ALPO achieves the best RQ and ES on both subsets, showing consistent benefits across the explicit--implicit distinction. ES differs little between the subsets (e.g., 52.23 vs.\ 51.60 for ALPO), indicating relatively stable Emotion Similarity under the evaluated conditions. We further neutralize vocal affect for 2,500 stratified test inputs by re-synthesizing each query with an all-zero emotion vector while preserving its transcript and speaker reference. IF drops by 39.52--41.08 points on implicit samples but 14.52--18.32 on explicit samples (Table~\ref{tab:prosody_intervention}), supporting that implicit intent relies more strongly on paralinguistic cues.

\begin{table}[t]
\centering
\small 
\begin{tabularx}{\columnwidth}{l *{8}{>{\centering\arraybackslash}X}}
\toprule
\multirow{2}{*}{Model} & \multicolumn{4}{c}{Explicit} & \multicolumn{4}{c}{Implicit} \\
\cmidrule(lr){2-5} \cmidrule(lr){6-9}
& IF & RQ & EA & ES & IF & RQ & EA & ES \\
\midrule
Base & 56.02  & 49.99  & 67.36   & 43.08 & 26.94 & 41.74 & 48.40 & 41.38 \\
SFT  & 93.50 & 84.54 & 89.60 & 43.37 & \underline{89.97} & \underline{83.49} & 86.68 & 42.62 \\
DPO  & 92.87 & 84.23 & \textbf{90.19} & 45.27 & 89.67 & 82.76 & \textbf{88.11} & 45.32 \\
GRPO & \underline{94.37}     & \underline{85.99} & 89.33       & \underline{49.76} & 89.59 & 83.22 & 85.37       & \underline{48.98} \\
ALPO & \textbf{94.50}     & \textbf{86.92} & \underline{89.77}         & \textbf{52.23} & \textbf{90.05} & \textbf{84.10} & \underline{87.67}          & \textbf{51.60} \\
\bottomrule
\end{tabularx}
\caption{Automatic evaluation results on the explicit and implicit subsets of the ParaIntent synthetic test set. }
\label{tab:explicit_implicit}
\end{table}

\begin{table}[t]
\centering
\small
\renewcommand{\arraystretch}{1.08}
\begin{tabularx}{\columnwidth}{l *{6}{>{\centering\arraybackslash}X}}
\toprule
Model
& $\mathrm{IF}_{\mathrm{Exp}}^{\mathrm{ori}}$
&  $\mathrm{IF}_{\mathrm{Exp}}^{\mathrm{neu}}$
& $\Delta\mathrm{IF}_{\mathrm{Exp}}$
&  $\mathrm{IF}_{\mathrm{Imp}}^{\mathrm{ori}}$
&  $\mathrm{IF}_{\mathrm{Imp}}^{\mathrm{neu}}$
& $\Delta\mathrm{IF}_{\mathrm{Imp}}$\\
\midrule
SFT  & 82.73 & 64.41 & 18.32 & 75.60 & 34.52 & 41.08 \\
GRPO & 83.64 & 69.08 & 14.56 & 77.24 & 37.08 & 40.16 \\
ALPO & 86.92 & 72.40 & 14.52 & 80.16 & 40.64 & 39.52 \\
\bottomrule
\end{tabularx}
\caption{Prosody neutralization results on 2,500 matched synthetic samples. IF ranges from 0 to 100, where $\Delta\mathrm{IF}=\mathrm{IF}^{\mathrm{ori}}-\mathrm{IF}^{\mathrm{neu}}$. All metrics use the same subset.}
\label{tab:prosody_intervention}
\end{table}

\subsection{Ablation Study}

Table~\ref{tab:ablation} compares independent per-reward advantage normalization without token-specific routing, following the normalization principle shared by GDPO and DVAO, with the full token-specific routing design under identical rewards.
Independent normalization
alone doesn't consistently improve over GRPO, whereas the
full ALPO design improves all four metrics. 
This indicates that
independent normalization is insufficient and that the gains
require reward-specific advantages to be applied to their
corresponding token subsets.

\begin{table}[t]
\centering
\small 
\begin{tabularx}{\columnwidth}{l *{6}{>{\centering\arraybackslash}X}}
\toprule
\multirow{2}{*}{Model} 
& \multirow{2}{*}{\shortstack{Indep.\\Adv.}} 
& \multirow{2}{*}{\shortstack{Adv.\\Routing}} 
& \multicolumn{2}{c}{Text} 
& \multicolumn{2}{c}{Speech} \\
\cmidrule(lr){4-5} \cmidrule(lr){6-7}
& & & IF & RQ & EA & ES \\
\midrule
GRPO                & $\times$ & $\times$ & \underline{91.99} & \underline{84.60} & 87.35 & \underline{49.37} \\
w/ Indep. Adv.      & \checkmark & $\times$ & 90.49 & 84.33 & \underline{87.75} & 49.29 \\
ALPO                & \checkmark & \checkmark & \textbf{92.28} & \textbf{85.51} & \textbf{88.72} & \textbf{51.92} \\
\bottomrule
\end{tabularx}
\caption{ALPO component ablation with identical rewards. \checkmark~and $\times$ indicate enabled and disabled components.}
\label{tab:ablation}
\end{table}

\subsection{Training Dynamics Analysis}
\label{sec:analysis}

Figure~\ref{fig:compare} compares GRPO and ALPO under identical reward functions. As shown in Figure~\ref{fig:compare} (a), the gradient norm ratio $\frac{\lVert g^{\mathrm{sp}}\rVert}{\lVert g^{\mathrm{txt}}\rVert}$ under standard GRPO remains below 1.0 throughout GRPO training, with a mean of 0.69, indicating weaker speech-side than text-side gradient magnitudes under a shared advantage. Under ALPO, the mean rises to 1.12, approaching balance between the token types. This shift is consistent with token-specific advantage routing producing more balanced text- and speech-side gradient magnitudes; it does not establish the absence of gradient conflict.

Figure~\ref{fig:compare} (b) shows the reward curves during training.
ALPO achieves higher final values on both $R^{\mathrm{sp}}$ and $R^{\mathrm{txt}}$ compared with GRPO, indicating that the two objectives benefit from independent advantage normalization and token-specific advantage routing rather than through a shared advantage. Together with Table~\ref{tab:ablation}, the results suggest that the benefit of ALPO primarily comes from routing each reward-specific advantage to its corresponding token type, allowing the textual and acoustic objectives to improve jointly.

\subsection{Discussion and Limitations}

RQ and ES are aligned with the training rewards and are
therefore interpreted alongside complementary task-level and
subjective evaluations. ParaIntent relies primarily on
generated text and synthesized speech, while its human-recorded
test set follows a controlled protocol rather than spontaneous
interaction. ALPO is evaluated on a single interleaved
text--speech backbone in a single-turn setting, leaving
multi-turn dialogue, broader architectures, and less
controlled real-world conditions for future work.

\section{Conclusion}
We introduced ParaIntent, a Chinese spoken emotional dialogue benchmark covering 14 intent categories with explicit and implicit expressions, together with a multidimensional evaluation protocol. We also proposed ALPO, which computes separate textual and acoustic advantages and routes them to the corresponding token subsets. Under matched rewards and training budgets, ALPO outperforms standard GRPO on most automatic metrics across both test sets and achieves the best subjective results among fine-tuned variants. The larger Intent Fulfillment drop after neutralizing vocal affect on implicit inputs further supports the explicit--implicit distinction. These results demonstrate the effectiveness of token-specific advantage routing for jointly improving response quality and emotional expression.

\bibliography{references}

\end{document}